\documentclass[useAMS,usenatbib]{mn2e}
\usepackage{psfig}

\def\msun{{\,M_\odot}}
\def\rsun{{\,R_\odot}}

\def\ledd{L_{\rm Edd}}

\newcommand\fe{Fe K$\alpha$\ }

\newcommand\sgra{Sgr~A$^{*}$ }

\newcommand\tchar{T_{\rm char}}

\def\>{$>$}
\def\<{$<$}

\def\simlt{\lower.5ex\hbox{$\; \buildrel < \over \sim \;$}}
\def\simgt{\lower.5ex\hbox{$\; \buildrel > \over \sim \;$}}
\def\sqr#1#2{{\vcenter{\hrule height.#2pt
      \hbox{\vrule width.#2pt height#1pt \kern#1pt
         \vrule width.#2pt}
      \hrule height.#2pt}}}

\def\del#1{{}}

\title[Inactive disk in \sgra] {Close stars and an inactive accretion
disk in Sgr~A$^*$: Eclipses and flares} \author[Nayakshin \& Sunyaev]
{Sergei Nayakshin$^1$ and Rashid Sunyaev$^{1,2}$ \\ $^1$
Max-Planck-Institut f\"ur Astrophysik, Karl-Schwarzschild-Str. 1,
85740 Garching, Germany\\ $^2$ Space Research Institute, Moscow,
Russia} \date{2003 Xxxxx XX}

\pagerange{\pageref{firstpage}--\pageref{lastpage}} \pubyear{2003}

\def\LaTeX{L\kern-.36em\raise.3ex\hbox{a}\kern-.15em
    T\kern-.1667em\lower.7ex\hbox{E}\kern-.125emX}

\begin{document}

\label{firstpage}

\maketitle

\begin{abstract}
A cold neutral and extremely dim accretion disk may be present as a
remnant of a past vigorous activity around the black hole in our
Galactic Center (GC). Here we discuss ways to detect such a disk
through its interaction with numerous stars present in the central
$0.1$ parsec of the Galaxy. The first major effect expected is X-ray
and near infrared (NIR) flares arising when stars pass through the
disk. The second is eclipses of the stars by the disk. We point out
conditions under which the properties of the expected X-ray flares are
similar to those recently discovered by Chandra. Since orbits of
bright stars are now being precisely measured (e.g. Sch\"odel et
al. 2002), the combination of the expected flares and eclipses offers
an invaluable tool for constraining the disk density, size, plane and
even direction of rotation. The winds of the O-type stars are
optically thick to free-free absorption in radio frequencies. If
present near \sgra core, such powerful stellar winds can modulate and
even occult the radio source.
\end{abstract}


\section{Introduction}\label{sec:intro}

The center of our Galaxy appears to host a very massive black hole
(e.g., Genzel 2000, Ghez et al. 2000) identified with the compact
radio source \sgra (e.g. Reid et al. 1999), with $M_{BH} \simeq 3
\times 10^6 \msun$. One mystery of \sgra is the fact that its
bolometric (mostly radio) luminosity is very low, i.e. $\sim 10^{-9}$
of the corresponding Eddington luminosity $\ledd = 1.3 \times 10^{44}
{\rm erg/sec} \; (M_{BH}/10^6 \msun)$ \, whereas its quiescent X-ray
emission is even dimmer than radio: $L_x\sim 10^{-11} \ledd$ (Baganoff
et al. 2001b; for a review of \sgra see Melia \& Falcke 2001 and
Markoff et al. 2001 on the role of the jet). This is puzzling because
there is enough hot gas observed at $\sim 0.04$ parsec from \sgra core
to let the black hole radiate some $\sim 4$ orders of magnitude more
(Baganoff et al. 2002a). The current favorite explanation of this
aspect of \sgra are accretion flow solutions (Narayan \& Yi 1994;
Narayan 2002) that radiate extremely little compared to the standard
disks (Shakura \& Sunyaev 1973).

However for ``our'' black hole to grow so massive there should have
been a much more vigorous accretion activity in the past. Such an
activity is usually assumed to proceed via the thin standard disk and
cease when the supply of matter ends. Note that practically
independently of the exact value of the specific angular momentum of
the accreting material, $a$, above the threshold value of $ \sqrt{3}
R_g c$, an accretion disk will develop over a broad range of radii
because matter in the disk flows in and out (Kolykhalov \& Sunyaev
1980). $R_g$ here is the gravitational radius, i.e., $R_g = 2 GM_{\rm
BH}/c^2 \simeq 9 \times 10^{11} $ cm. A ``light'' and very cold
($T\sim 10^2-10^3$ K) inactive disk may remain there essentially
indefinitely because its viscosity is extremely low. Nayakshin (2002)
recently suggested that there is such a disk in \sgra and that it is
draining the heat from the hot gas by thermal conduction. Instead of
flowing into the black hole the hot gas in this model settles down
(condensates) onto the inactive disk at large distances. The accretion
of the hot gas is thus delayed in this picture which then would
explain the low luminosity of Sgr~A$^*$.

\sgra is believed to be closely related to the Low Luminosity AGN
(LLAGN; e.g. Ho 1999). Most if not all of these sources seem to have
cold disks that often can be seen only through water maser emission
(e.g. Miyoshi et al. 1995) arising in a range of radii where gas
temperature is $200-1000$ K (Neufeld \& Maloney 1995). These disks
also seem to miss their innermost parts (Quataert et
al. 1999). Unfortunately our GC is an extremely ``low luminosity''
LLAGN and an inactive disk can be very hard to detect. In particular,
a thermally emitting inactive accretion disk with $T_d\simlt 100$ K
and $R_d\la\; \hbox{few}\; 10^4 R_g$ is dim enough not to violate the
quiescent bolometric luminosity constraints. In addition, disks of
such low temperature are razor-thin, i.e. $H/R\sim 10^{-3}$.  If the
disk happened to be oriented nearly edge-on to us, it would be very
difficult to spot it via its {\em quiescent} emission.

However there seems to be as many as $\sim 10^4$ stars in the central
arcsecond ($1\arcsec \simeq 0.039$ pc $\simeq 1.2\times 10^{17}$ cm)
of the Galaxy. The orbits of brightest of these stars (B and possibly
O type; Gezari et al. 2002) are now being precisely mapped
(e.g. Sch\"odel et al. 2002). These stars may be eclipsed by the disk
if it is optically thick. In addition, even much smaller and much less
luminous solar-type stars will produce X-ray and near infra-red flares
when passing through the disk. In this Letter we hope to present a
concise but clear overview of our ongoing work on the star-disk
interactions near GC (Nayakshin, Cuadra \& Sunyaev 2003, in
preparation).

\del{Our main conclusion is that close resemblance of predicted X-ray
flares to those observed by Chandra is supportive of the disk idea,
but}

\section{Star-disk eclipses and crossings}\label{sec:eclipses}

\begin{figure}
\centerline{\psfig{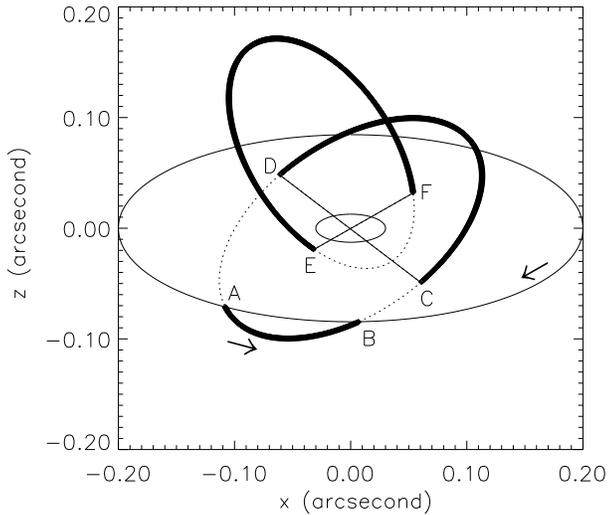}}
\caption{Two examples of star-disk eclipses and flares. Visible
and eclipsed parts of stars' trajectories are shown with thick solid
and dotted curves, respectively.}
\label{fig:fig1}
\end{figure}
\begin{figure}
\centerline{\psfig{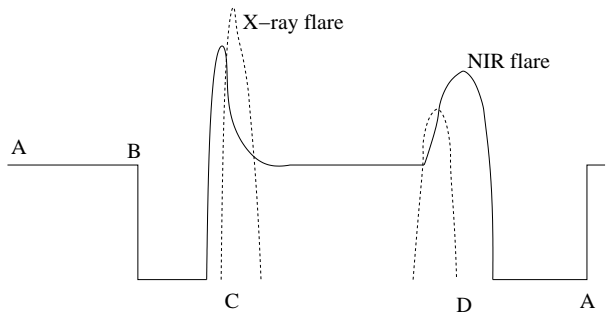}}
\caption{Schematic light curve of the star-disk eclipses and flares
corresponding to the orbit ABCD in Figure \ref{fig:fig1}. Flares,
shown not to scale, are shorter by about a factor of $10^3$ than
eclipses. Their duration and amplitude can be used to
constrain disk properties.}
\label{fig:fig2}
\end{figure}

A star can be eclipsed when it moves into the area shadowed by the
disk as seen from our line of sight. If the star moves ``far''
behind the disk then this type of eclipses is completely analogous to
the eclipses of Sun by the Moon and can be referred to as a ``blocking
eclipse''. As such it happens because both the blocked and the
blocking objects have finite sizes. The star may instead strike the
disk, pass through it, and become eclipsed because it moved from the
front/visible side of the disk to its back/invisible side. This type
of eclipses may be referred to as an ``impact eclipse''.

\begin{figure}
\centerline{\psfig{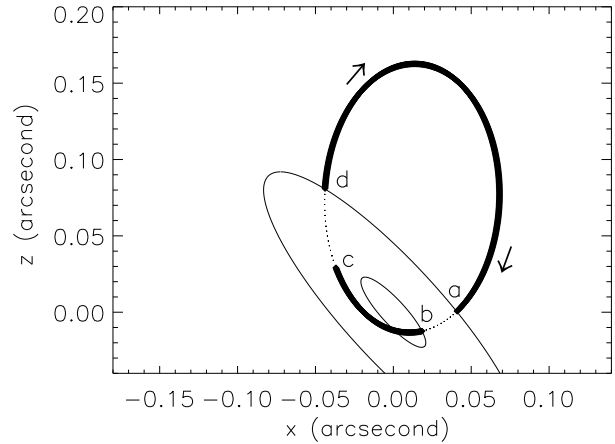}}
\caption{Eclipses and a flare for star S2 (Sch\"odel et al. 2002) and
a disk inclined at $i=75^\circ$ with $R_d=0.12$\arcsec ($\sim 1.4
\times 10^{16}$ cm) and inner radius $R_i=0.03$\arcsec. Highly
elongated orbits are excellent probes of the inner disk structure.}
\label{fig:fig3}
\end{figure}
\begin{figure}
\centerline{\psfig{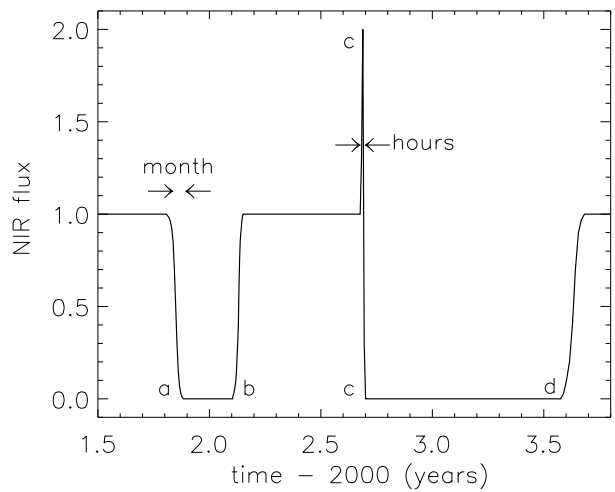}}
\caption{Lightcurve for the star-disk system shown in
Fig. \ref{fig:fig3}.  Note the large difference in time scales of
flares and duration of eclipses.}
\label{fig:fig4}
\end{figure}

In Figure (\ref{fig:fig1}) we show two star trajectories interesting
in terms of eclipses and flares. The star moving along the ABCD path
is eclipsed by the disk twice per orbit. In point B the star does not
physically intercepts the disk but only its shadow, and hence this is
the beginning of a blocking eclipse. Due to the star's finite size and
(most likely) a gradual rather than an abrupt nature of the disk edge
on both the inner and the outer boundaries, partial eclipses could be
observed for hours to months (see Fig. \ref{fig:fig4}).  The eclipse
ends at point C when the star goes through the disk, from its back to
its front side. NIR infra-red and an X-ray flare will be emitted (see
Figure \ref{fig:fig2}). {\em If the disk is optically thick} to
X-rays, then NIR flare will precede the X-ray flare. The former is
thermal in nature and is strongest when the star is in the disk
midplane, whereas the latter can only reach the observer when the star
is in the disk photosphere (see \S \ref{sec:scheme}). The CD part of
the trajectory is an ``unremarkable'' quiescent star emission, which
terminates when the star hits the disk for the second time. This time
the X-ray flare should be emitted first. The eclipsed part DA ends
when the star emerges from the disk shadow without a flare.

To date, there is only one star whose orbit is nearly completely
known, i.e. the star S2 (Sch\"odel et al. 2002). The orbit is highly
elliptical (eccentricity $e=0.87$), inclined at $i_* = \pm 46^\circ$
(we picked the $+$ sign here), and the period is 15.2 years. We plot
(Figure \ref{fig:fig3}) the orbit of S2 and a disk that does {\em not}
eclipse any of the measured positions (Fig. 1 in Sch\"odel et
al. 2002). Note that the disk can in principle still exist beyond
$R_i$ or $R_d$ but it cannot be optically thick there.  Figure
(\ref{fig:fig4}) shows the expected lightcurve of the star (the
passage of the periblackhole radius for S2 occured at 2002.3). The
eclipse $ab$ ($t\sim 2002.0$) is of the blocking type and therefore
there is no flare associated with it, whereas the second eclipse
(point $c$) begins with a flare. Figures (\ref{fig:fig3} \&
\ref{fig:fig4}) are only examples. We find that a disk with no inner
hole or a disk with a larger inner hole, $R_i \simgt 0.05$\arcsec and
an arbitrarily large outer radius are also possible for the present
data (Cuadra et al. 2003, in preparation).

A brief summary of what one can learn from observations of star
eclipses and flares:
\begin{enumerate}
\item Coordinates of star's trajectories are three-dimensional;
therefore, star-disk crossing points (e.g. D or C) yield 3D coordinate
of a point in the disk. Together with the known coordinates of \sgra
itself, any such crossing yields a line in the disk plane. Therefore,
knowledge of 3D coordinates of points D \& F, for example, (and
Sgr~$A^*$) is sufficient to determine the plane of the disk rotation.

\item If the plane of the disk is known, then the projected
coordinates of two blocking points (A \& B) uniquely determine the
outer disk radius (unless the disk is strongly warped).

\item Appearance or disappearance of stars (in blocking points) will
be gradual and this should yield information on how the disk is
terminated on both boundaries.

\item The maximum luminosity of a flare is a strong function of the
angle between the star and the disk velocity at the point of impact
(\S \ref{sec:scheme}), $\theta_r$. Hence one can tell
whether the disk is rotating clockwise or counter-clockwise in
Fig. (\ref{fig:fig1}).  The best constraints here are offered by a
star whose orbital plane is close to the disk plane.

\item In general, a star that is on an elliptical orbit and that
strikes the disk twice per orbit, will do so at different radii from
the black hole (e.g. F is farther from the center than E in
Fig. \ref{fig:fig1}), so the flare durations, amplitudes and spectra
will be different. Such events will offer better constraints on the
variation of the disk density and temperature with $R$ than will
flares from separate stars.
 


\item Currently orbits of only the bright stars can be followed in
\sgra star cluster. These stars have radii $R_* \sim 5-10 \rsun$ and
hence should yield very large X-ray and NIR flares (see \S
\ref{sec:scheme}). Of course such flares are much less frequent than
the typical flares from stars with $R_*\sim \rsun$.
\end{enumerate}

\section{Physics of star-disk flares}\label{sec:scheme}

Reader not interested in physical or mathematical detail of our model
for the star-disk flares may proceed directly to \S 3.1 and Figure
(\ref{fig:fig5}) where star-disk flare properties are summarized.

At a distance $R$ from the black hole, let a star with mass $M_* = m_*
\msun$ and radius $R_* = r_* \rsun$ move with velocity $v_*$ that
makes angle $\theta_r$ with the disk circular Keplerian velocity,
$v_K$. We also define for convenience $r_4 = R/10^4 R_g$.  The
relative velocity of the disk and the star, $v_{\rm rel}$ is essential
for the problem. In general $v^2_{\rm rel} = v_K^2 + v_*^2 - 2 v_* v_K
\cos\theta_r$, and one needs to carry out calculations for arbitrary
$v_*$ and $\theta_r$ and then integrate over the star's 3D velocity
distribution. In what follows we simply assume that $v_* \simeq
v_K$. Under the assumption of an isotropic star cluster the
``average'' angle $\theta_r$ is equal to $ \pi/2$, so $v _{\rm rel} =
\sqrt{2} v_K$. However we shall keep in mind the importance of the
actual $v_*$ and $\theta_r$ for exact results (cf. point iv above).

The star's internal density is very much larger than that of the disk
and we can consider the star to be a rigid solid body (see Syer et
al. 1991). We neglect tidal effects and accretion of gas onto the
stellar surface because the corresponding Bondi radius is much smaller
than $R_*$ ($R_{\rm B} = GM_*/v_{\rm rel}^2 = 1.5 \times 10^9 m_* r_4
\ll 7 \times 10^{10} r_*$).  The Mach number of the star in the disk
is about $v_{\rm rel}/\sqrt{(kT_{d}/m_p)}\sim 10^4$.  Thus the star
drills a narrow hole in the disk (Syer et al. 1991) Note also that
$R_*\ll H$. The star is essentially a piston moving into the gas and
the rate at which the star makes work on the gas in the disk is
approximately $L_w \sim \pi R_*^2 m_p n_d v_{\rm rel}^3$, where $n_d$
is the midplane density of hydrogen nuclei.

\paragraph*{X-ray spectrum.} 
The characteristic temperature to which the gas is heated in the
shock wave is
\begin{equation}
\tchar = \frac{2}{3} \frac{\mu v_{\rm rel}^2}{2 k} = 1.8 \times
10^{8} r_4^{-1}\; {\rm K},
\label{tchar}
\end{equation}
where we set $\mu \simeq m_p/2$. 
For $\tchar$ as high as $10^8$ K, optically thin X-ray spectra are
dominated by bremsstrahlung emission. The photon spectral index in the
$2 \simlt E \simlt 8$ keV energy range is $\Gamma \simlt 1.5$ and is
in agreement with the observed values of $\Gamma\simeq 1\pm 0.7$
(Baganoff et al. 2001, Goldwurm et al. 2002). \fe line emission is
weak as observed. Absorption of soft X-rays in the disk photosphere
could be significant. However we find that X-rays from {\em bright}
flares photo-ionize the disk material enough to hide this absorption
on the background of the existing neutral absorber in the line of
sight.

\paragraph*{X-ray luminosity.}

Consider first an optically thin case.  If the cooling time of the
shocked gas, $t_c$, is shorter than $R_*/v_*$, then the X-ray
luminosity, $L_{\rm xe}$, should be about $L_w$. On the other hand, if
adiabatic losses dominate cooling of the hot gas, then $L_{\rm xe}\sim
L_w (R_*/v_*t_c)\ll L_w$. We find that the latter case is more
appropriate for \sgra flares, for which we obtain
\begin{equation}
L_{\rm xe} \sim 4.1 \times 10^{33} \; n_{11}^2 r_*^3 r_4^{-1/2} \;
{\rm erg/sec}\;,
\label{lxe}
\end{equation}
where $n_{11} = n_d/10^{11}$ cm$^{-3}$.

If the disk is optically thick to photo-absorption, then X-ray
emission becomes observable only when the star reaches the disk
photosphere, i.e.  when $\tau = (\cos{i})^{-1} \sigma_{\rm eff} H n_p
\simeq 1$, where $i$ is the disk inclination angle, $H$ is disk
half-thickness, $\sigma_{\rm eff} = b \sigma_T $ is the effective
X-ray total cross section ($b \sim$ few), and $n_p$ is local
density. This moment defines the maximum in the X-ray lightcurve;
inside the disk X-rays cannot escape to the observer; in the disk
photosphere, on the other hand, the gas density is much smaller than
it is in the midplane.

Using the $\tau=1$ condition we then obtain the density $n_{\rm max}
\sim \cos{i}/(b \sigma_T H)$ at which the maximum X-ray luminosity,
$L_{\rm max}$, is emitted. $L_{\rm max}$ is clearly given by the
optically thin luminosity (eq. \ref{lxe}), calculated with $n_{\rm
max}$ instead of $n_d$, times $\exp[-1]$:
\begin{equation}
L_{\rm max} \sim 2 \times 10^{34}\; \left[\frac{\cos{i}}{b}
\right]^2 T_2^{-1} r_*^3 r_4^{-7/2} \; \hbox{erg/sec}\;.
\label{lmax}
\end{equation}

\paragraph*{Flare duration.}

The disk half thickness, $H$, is
\begin{equation}
H = \sqrt{\frac{k T_d R^3}{G M_{\rm BH} m_p}}= 3.9 \times 10^{12} \,
T_2^{1/2} r_4^{3/2} \; {\rm cm}\;,
\label{h}
\end{equation}
where $T_2 = T_d/100$ K, and we assumed that the gas is mainly
molecular hydrogen. The flare duration for an optically thick disk is
$t_{\rm dur} \sim (H + 2 R_*)/v_*$:
\begin{equation}
t_{\rm dur} \simeq 2.5 \times 10^4 \; T_2^{1/2} r_4^2 + 670\; r_*
r_4^{1/2} \; {\rm sec}\;,
\label{tdur}
\end{equation}
which is in accord with observations if $R\sim 10^3 - 10^4
R_g$.

\paragraph*{Mass of the disk.} 
The disk surface density $\Sigma \simeq 2 H n_d m_p = 12.8 \; n_{12}
r_4^{3/2}\, T_2^{1/2}$ g cm$^{-2}$. The mass of the accretion disk,
calculated for density and temperature independent of radius, is $M_d
\sim \msun \, n_{12} r_4^{7/2} T_2^{1/2}$.  Such a ``light'' disk is
neither globally nor locally self-gravitating (e.g. Kolykhalov \&
Sunyaev 1980; Gammie 2001).

\del{Results of Pollack et al. (1994) show that a dusty accretion disk at
$T_2\sim 1$ would be optically thick for $\Sigma \sim 1$ so we assume
that deep in an optically thick disk, the X-rays will be absorbed and
re-emitted as a black body emission in the near infrared (NIR).}

\paragraph*{NIR Luminosity.}

Consider optically thick case, $\Sigma \ga 1$ g/cm$^2$. Deep inside
the disk X-rays are absorbed and re-emitted as a black body emission
in the near infrared.  Assuming $L_w \sim \sigma_B T_{ir}^4 \times 2
(2 H)^2$, we get $T_{ir} \simeq 2.7 \times 10^3 \;{\rm K}\;
[n_{12}r_*^2/T_2]^{1/4} r_4^{-9/8}$.  The predicted NIR luminosity
is then $\nu L_{\nu} = 4 H^2 \cos{i}\; \nu B_{\nu}(T_{ir})$:
\begin{equation}
\nu L_{\nu} = 4.5 \times 10^{35} \cos{i}\; T_2\, r_4^3\;
\frac{\bar{\nu}^4}{e^x - 1} \quad \hbox{erg/sec}\;,
\label{lnir}
\end{equation}
where $\bar{\nu}\equiv \nu/1.5 \times 10^{14}$, and $x\equiv
h\nu/kT_{ir}$; we assumed the distance to the GC $D=8.0$ kpc. While
the luminosity in the optically thin case will be different from that
given in eq. (\ref{lnir}), we continue to use the latter as a rough
guide in both cases.

\paragraph*{Number of star-disk flares per year.}

A cusp with stellar density $n_*(R) = n_{0*} (R/R_c)^{-p}$, with $p =
7/4$ is predicted by Bahcall \& Wolf (1976) for $R_t < R < R_c$, where
$R_t \simeq 20 R_g m_*^{-1/3}$ is the tidal radius and $R_c$ is the
cusp radius.  Alexander (1999) finds that $n_*(R)$ with $p$ ranging
from $3/2$ to $7/4$ is {\em slightly} favored over a constant density
one (i.e. $p=0$). Eckart et al. (2003) finds that the mass of a
putative cusp is $M_{\rm cusp}\sim 5\times 10^3 \msun$ and $R_c\simeq
0.14\arcsec \simeq 1.7 \times 10^{16}$ cm. Thus $n_{0*} \simeq 3.8
\times 10^9\; m_*^{-1}$ pc$^{-3}$. As an example, we give the number
of star-disk crossings per year for $p=3/2$:
\begin{equation}
\dot{N}(R) \simeq 6.7 \times 10^3 \; m_*^{-1}\; \frac{\ln(R_d/R_i)
}{\ln(500) } \; \hbox{year}^{-1}\;,
\label{er}
\end{equation}
where we assumed $R_i > R_t$. This is the rate of {\em star-disk
crossings}, and not all of these will produce flares strong enough to
be detected. Within current uncertainties
the expected rate of flares compares quite well with the observed
$\sim 1$ flare per day (Baganoff et al. 2002b) for power-law indices
$p$ ranging from $\sim 1$ to $\sim 2$, but would be too low if $p\sim
0$ unless $M_{\rm cusp}$ is greatly under-estimated.

\paragraph*{Inner hole in the disk.} 

The star-disk interactions can remove the disk angular momentum and
yield some accretion. The time to remove all disk angular momentum
(found as in Ostriker 1983) is shorter than $10^6$ years for radii
less than $\sim 2 \times 10^3 R_g$ where we assumed $p=3/2$.
The disk viscous time at $r_3 = R/10^3 R_g$ is $t_{\rm visc}\sim 5
\times 10^5 \alpha^{-1} T_2^{-1} r_3^{1/2}$ years, where $\alpha <
1$ is viscosity parameter (Shakura \& Sunyaev 1973). Clearly the disk
inner hole cannot be refilled by accretion of material from larger
radii for $R \la 10^3 R_g$. In addition, any ``minor'' accretion
events in the past would heat up the inner disk and destroy it.

\subsection{Summary of flare properties}

We plot flare properties in Figure (\ref{fig:fig5}) for two values of
the star's radius, $R_* = \rsun$ (thin curves) and $R_*=5 \rsun$
(thick).  Conclusions from Figure (\ref{fig:fig5}):

\begin{figure}
\centerline{\psfig{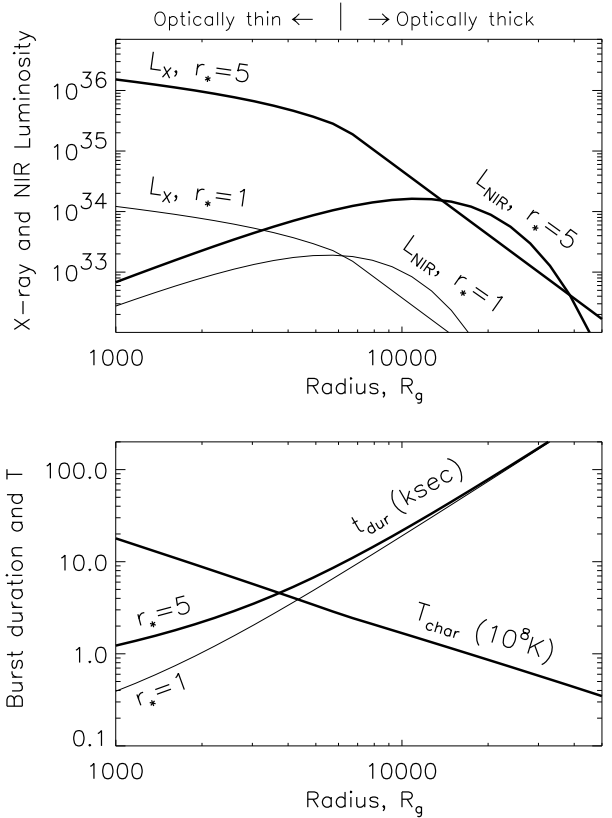}}
\caption{Dependence of flare characteristics on the radial distance
from the black hole. The disk has a constant midplane density $n_d =
10^{11}$ cm$^{-3}$ and is inclined at $i=75^\circ$. The disk region
with $R\la 7 \times 10^{3} R_g$ is optically thin along the line of
sight whereas for $R\simgt 7 \times 10^{3} R_g$ it is optically thick.
Two values for the star radius, $r_*=R_*/\rsun$ are considered. $b=2$
for all the curves. {\bf Top panel:} maximum X-ray and near infrared
(NIR) luminosities achieved during bursts. Note the break at the slope
of $L_x$ where the disk becomes optically thin. Star star-disk
collisions at $R\simgt \;{\rm few} \times 10^4 R_g$ produce X-ray
flares too weak to observe.  {\bf Lower panel:} Burst duration and
characteristic temperature.}
\label{fig:fig5}
\end{figure}

\begin{enumerate}

\item Flare X-ray luminosity is a strong function of radius
$R$. Flares even from large stars are too dim to be detected for
$R\simgt 2 \times 10^4 R_g$. Strong flares are produced by large ($R_*
> \rsun$) stars. If disk is optically thick, X-ray spectra of weak
flares should be influenced by photo-absorption in the disk
photosphere. 

\item To date no flares from \sgra much in excess of $L_x \sim
10^{35}$ erg/sec were observed by Chandra or XMM. This requires the
disk midplane density be $n_d\simlt 10^{11}$ cm$^{-3}$. The infrared
opacity of dust is $k_{nir} \sim \; \hbox{few}$ cm$^2$/gm (Fig 3b in
Pollack et al. 1994).  The disk is optically thin in NIR for $R\la\;
\hbox{few}\; \times 10^3 R_g$.

\item The disk cannot be exactly edge-on to us or else X-ray flares
would be too weak (see eq. \ref{lmax}). 


\item The maximum NIR 2 $\mu$m luminosity for solar-type stars is
$\sim \;\hbox{few}\;\times 10^{34}$ erg/sec. Such flares cannot be
detected because the current limit on NIR flares (Hornstein et
al. 2002) is about $20$ mJy or $2\times 10^{35}$ erg/sec.  Further,
the flares are actually offset from \sgra location by $\sim$
0.1\arcsec. If detected they would probably appear as ``noise'' not
associated with \sgra itself.

\item Similar X-ray and NIR flares from star-disk interactions in more
distant LLAGN with $M_{\rm BH} \gg 3\times 10^6 \msun$ have smaller
luminosities (typically $n_d \propto 1/M_{\rm BH}$).  In accord with
this, no such X-ray flares have yet been observed in other LLAGN. 

\item Radio emission from the shocked gas is very weak (compared to
the NIR or X-ray luminosities) since it is completely self-absorbed.
This explains why there is no radio flares correlated with X-ray
outbursts in Sgr~A$^*$.

\item The bolometric luminosities of the O-type stars can be as high
as $10^{39}$ erg/sec. Passing through the disk their radiation will
form a strong ionization front and will be reprocessed into a cooler
thermal emission. The emission will be coming from a larger surface
area than $\pi R_*^2$. The flare NIR K-band luminosity should be much
larger than the star quiescent emission (in the same band).

\end{enumerate}

\section{Modulation of jet emission by stars}

The luminous stars produce powerful winds with mass loss rates as high
as $\dot{M}_w = 10^{-5} \msun$ per year.  The optical depth to
free-free absorption in radio frequencies, $\tau_{\rm ff}(\nu)$, can
be significant. Estimating density in the wind a distance $\tilde{R}$
from the star as $n_H \sim \dot{M}_w/4\pi \tilde{R}^2 v_w m_p$, we
obtain
\begin{equation}
\tau_{\rm ff} \sim 2\; \dot{M}_6^2 v_3^{-2} T_4^{-3/2}
[\nu/10^{11}\;\hbox{Hz}]^{-2}\; [\tilde{R}/10^{13}
\;\hbox{cm}]^{-3}\;,
\label{tff}
\end{equation}
where $T_4 $ is wind temperature in units of $10^4$ K, and $v_3\equiv
v_w/10^3$ km/sec, and $\dot{M}_6=\dot{M}_w/10^{-6} \msun$ per
year. Therefore such powerful stellar winds may eclipse jet radio
emission at low enough frequencies. Unfortunately probability of this
event is extremely low because there are very few (if any!) of such
stars within small enough distance from Sgr~A$^*$. 

A much more likely effect is a star passing through the jet. The jet
emission can be either enhanced (via providing seed photons from the
star) or reduced (via providing too many seed photons which over-cool
the jet electrons). The rate of such ``jet-crossing'' events is
roughly $\theta_j/\pi$ times smaller than that for the star-disk
crossings, where $\theta_j$ is the jet opening angle. For example, for
$\theta_j=3^\circ$ the rate is 1/60 of that given by eq. (\ref{er}),
i.e. $\sim 100$ year$^{-1}$.  However since \sgra radio power is much
greater than its quiescent X-ray emission only the brightest stars are
likely to cause jet disturbance. Therefore one may expect probably few
events per year. The duration of the event is much longer than that of
X-ray flares, i.e. $t_{\rm dur} \sim \theta_j R/v_*$ i.e. weeks to
years depending on the distance $R$.  Interestingly, Zhao et
al. (2001) recently reported discovery of a 106 day periodicity in the
radio emission of Sgr~A$^*$.

\section{Discussion}\label{sec:discussion}

We have shown that eclipses of close stars by the disk may be a very
effective tool with which to constrain the disk's plane and direction
of rotation, inner and outer radii, and the optical depth variation
with radius. The predicted X-ray and NIR flares may yield additional
constraints on the properties of the disk.  The disk density is the
only parameter of the model that is completely free; all the others --
the star's radii, number density, disk size, temperature and
inclination angle have certain limitations from observations of either
stellar orbits near \sgra or its quiescent emission. It is thus
remarkable that a disk with midplane density $n_d\sim \; \hbox{few}\;
10^{11}$ cm$^{-3}$ produces X-ray flares with ``right'' duration,
X-ray luminosity, spectrum, and low enough NIR and radio emission to
pass all of the observational constraints on \sgra flares. If the
inactive disk indeed exists, it is necessary to understand how it
 interacts with the hot (ADAF-type) flow.

\del{Chandra has now resolved the source of the hot gas presumably feeding
the black hole in Sgr~A$^*$. If the inactive disk indeed exists, it is
necessary to understand how the cold disk and the hot flow
interact. It would be especially important to understand whether the
hot gas condenses onto the cold disk (delayed accretion) or the hot
gas evaporates the cold disk, or there is simply no significant
interaction between the two.}

\section{ACKNOWLEDGMENTS}

Useful discussions with P. Predehl, R. Schoedel, T. Ott, and
especially Jorge Cuadra are greatly appreciated.

{}

\label{lastpage}


\begin{thebibliography}{}

\bibitem[]{} Alexander, T. 1999, ApJ, 527, 835

\bibitem[\protect\citeauthoryear{Baganoff et~al.}{Baganoff
et~al.}{2001}]{Baganoffetal2001} Baganoff F.~K., et~al., 2001, Nature,
413, 45

\bibitem[\protect\citeauthoryear{Baganoff et~al.}{Baganoff
et~al.}{2001b}]{Baganoffetal2001b} Baganoff F.~K., et~al., 2002a, ApJ,
in press

\bibitem[\protect\citeauthoryear{Baganoff et~al.}{Baganoff
et~al.}{2002b}]{Baganoffetal2002b} Baganoff F.~K., et~al., 2002b, in
the GC workshop ''The central 300 parsecs'', Hawaii,
November 2002.

\bibitem[]{} Bahcall, J.N., \& Wolf, R.A. 1976, ApJ, 209, 214

\bibitem[]{} Eckart, A., in the GC workshop ''The central 300
parsecs'', Hawaii, November 2002.

\bibitem[]{} Gammie, C.F. 2001, ApJ, 553, 174

\bibitem[]{} Gezari, S., et al. 2002, ApJ, 576, 790

\bibitem[]{} Ghez, A., et al. 2000, Nature, 407, Issue 6802, 349


\bibitem[\protect\citeauthoryear{{Goldwurm} et~al.}{{Goldwurm}
et~al.}{2002}] {Goldwurmetal2002}{Goldwurm}, A., et al. 2002, ApJ in
press, astro-ph/0207620

\bibitem[]{} Ho, L. 1999, ApJ, 516, 672

\bibitem[]{} Hornstein, S.D. et al. 2002, ApJ, 577, L9

\bibitem[]{} Kolykhalov, P.I., \& Sunyaev, R.A. 1980,
Sov. Astron. Lett., 6, 357




\bibitem[\protect\citeauthoryear{{Markoff} et~al.}{{Markoff}
et~al.}{2001}]{Markoffetal2001} {Markoff} S., et al. 2001, A\&A, 379,
L13

\bibitem[]{} Melia, F., \& Falcke, H. 2001, ARA\&A, 39, 309

\bibitem[]{} Miyoshi, M., et al. 1995, Nature, 373, 127

\bibitem[]{} Narayan, R., \& Yi, I. 1994, ApJ, 428, L13


\bibitem[\protect\citeauthoryear{{Narayan}}{Narayan}{2002}]{Narayan2002}
{Narayan} R., 2002, ``Lighthouses of the Universe'', Proceedings of
the MPA/ESO/, p. 405

\bibitem[]{} Nayakshin, S. 2002, in the GC workshop ''The central 300
parsecs'', Hawaii, November 2002.


\bibitem[]{} Neufeld, D.A., \& Maloney, P.R. 1995, ApJL, 447, L17


\bibitem[]{} Ostriker, J.P. 1983, ApJ, 273, 99

\bibitem[]{} Pollack, J.B., et al. 1994, ApJ, 421, 615



\bibitem[]{} Quataert, E., et al. 1999, ApJL, 525, L89

\bibitem[\protect\citeauthoryear{{Reid} et~al.}{{Reid}
et~al.}{1999}]{ReidReadheadVermeulen1999} {Reid} M.~J., et al. 1999,
ApJ, 524, 816


\bibitem[\protect\citeauthoryear{{Sch\"odel} et~al.}{{Sch\"odel}
et~al.}{2002}]{Schodel2002} {Sch\"odel} R., et al. 2002, Nature, 419,
694

\bibitem[]{} Shakura, N.I., \& Sunyaev, R.A. 1973, A\&A, 24, 337





\bibitem[\protect\citeauthoryear{{Zhao}, {Bower}, \& {Goss}}{{Zhao}
  et~al.}{2001}]{ZhaoBowerGoss2001}
{Zhao} J., {Bower} G.~C.,  {Goss} W.~M., 2001, ApJL, 547, L29

\end{thebibliography}
\end{document}